\documentclass[11pt, twocolumn]{article}

\usepackage{caption}
\usepackage{acl}
\usepackage{natbib}
\usepackage{times}
\usepackage{amsmath}
\usepackage{amssymb}
\usepackage{graphicx}
\usepackage{booktabs}
\usepackage{hyperref}
\usepackage{geometry}
\usepackage{enumitem}
\usepackage{xspace}
\usepackage{float}
\usepackage{tcolorbox}
\usepackage{subcaption}
\tcbuselibrary{breakable}
\usepackage{tikz}
\usetikzlibrary{arrows.meta, positioning, shapes.geometric}
\usepackage{pgfplots}
\pgfplotsset{compat=1.18}

\title{\bfseries
Scaling Retrieval Augmented Generation with RAG Fusion: Lessons from an Industry Deployment
}

\author{
Luigi Medrano \\
Dell Technologies \\
Austin, TX, USA \\
{\small \texttt{Luigi.Medrano@dell.com}}
\And
Arush Verma \\
Dell Technologies \\
Austin, TX, USA \\
{\small \texttt{Arush.Verma@dell.com}}
\And
Mukul Chhabra \\
Dell Technologies \\
Austin, TX, USA \\
{\small \texttt{mukul.chhabra@dell.com}}
}

\date{}

\renewcommand{\abstractname}{\centering\textbf{Abstract}}

\begin{document}
\maketitle

\begin{abstract}
Retrieval-Augmented Generation (RAG) systems commonly adopt retrieval fusion techniques such as multi-query retrieval and reciprocal rank fusion (RRF) to increase document recall, under the assumption that higher recall leads to better answer quality. While these methods show consistent gains in isolated retrieval benchmarks, their effectiveness under realistic production constraints remains underexplored. In this work, we evaluate retrieval fusion in a production-style RAG pipeline operating over an enterprise knowledge base, with fixed retrieval depth, re-ranking budgets, and latency constraints.

Across multiple fusion configurations, we find that retrieval fusion does increase raw recall, but these gains are largely neutralized after re-ranking and truncation. In our setting, fusion variants fail to outperform single-query baselines on KB-level Top-$k$ accuracy, with Hit@10 decreasing from $0.51$ to $0.48$ in several configurations. Moreover, fusion introduces additional latency overhead due to query rewriting and larger candidate sets, without corresponding improvements in downstream effectiveness.

Our analysis suggests that recall-oriented fusion techniques exhibit diminishing returns once realistic re-ranking limits and context budgets are applied. We conclude that retrieval-level improvements do not reliably translate into end-to-end gains in production RAG systems, and argue for evaluation frameworks that jointly consider retrieval quality, system efficiency, and downstream impact.
\end{abstract}

\section{Introduction}

Retrieval-Augmented Generation (RAG) systems have emerged as a widely adopted approach for grounding large language models (LLMs) in external knowledge bases~\citep{lewis2020rag}. By conditioning generation on retrieved documents, RAG systems aim to improve factual accuracy and reduce hallucinations. However, their effectiveness remains highly dependent on the quality of retrieval: when relevant evidence is missing or poorly ranked, downstream generation is constrained regardless of model capacity.

A common strategy for improving retrieval robustness is the use of retrieval fusion techniques, such as multi-query retrieval and reciprocal rank fusion (RRF) \citep{cormack2009rrf}. By issuing multiple reformulations of a query or combining results from different retrieval passes, fusion methods aim to increase document recall and reduce missed evidence. Prior work has shown that these techniques can improve recall-oriented metrics in isolation, motivating their adoption in many RAG pipelines.

Despite their popularity, the end-to-end impact of retrieval fusion in production RAG systems remains unclear. In practice, retrieval operates under additional constraints, including fixed retrieval depth, limited re-ranking capacity, context window budgets, and strict latency requirements. Under these conditions, increases in raw recall may be offset by redundancy, conflicting context, or truncation effects introduced downstream.

In this paper, we investigate whether retrieval fusion improves system-level performance in a production-style RAG setting. Rather than proposing new fusion algorithms, our focus is on evaluating whether recall gains observed at the retrieval layer translate into meaningful improvements after re-ranking and context selection. Our results show that while fusion increases candidate diversity, these gains are often neutralized once realistic pipeline constraints are applied, resulting in limited or negative impact on KB-level accuracy and latency.

\subsection{Contributions}

This paper makes the following contributions:

\begin{itemize}
    \item We present a production-oriented evaluation of retrieval fusion methods in a RAG pipeline operating under fixed retrieval depth, re-ranking budgets, and latency constraints.
    \item We empirically analyze the relationship between retrieval-level recall gains and downstream KB-level accuracy, showing that fusion-induced recall improvements are frequently neutralized after re-ranking and truncation.
    \item We identify practical trade-offs introduced by fusion, including increased redundancy and additional latency overhead, that limit its effectiveness in realistic RAG deployments.
\end{itemize}

\section{Background}

Retrieval-Augmented Generation (RAG) systems were adopted in this work to support factual question answering over large, public facing enterprise knowledge bases, where standalone large language models (LLMs) are insufficient. By conditioning generation on retrieved evidence, RAG enables grounded, traceable responses under operational constraints such as access control, latency, and evolving content.

In practice, early deployments revealed that end-to-end performance was often limited by retrieval rather than generation. Short, underspecified, or weakly lexicalized queries—common in real user interactions—frequently failed to surface the correct knowledge-base articles, even when using strong hybrid retrievers. These retrieval failures propagated downstream, constraining the generator’s output and resulting in incomplete or low-quality responses despite capable LLMs.

To address these recall failures, retrieval fusion techniques such as multi-query retrieval and Reciprocal Rank Fusion (RRF) have been proposed to increase document coverage by aggregating results from multiple retrieval signals ~\citep{rackauckas2024ragfusion}. Prior work reports consistent improvements on recall-oriented metrics, particularly for ambiguous or short queries, motivating the adoption of fusion in modern RAG pipelines.

However, in production systems, retrieved candidates are subject to reranking, deduplication, and truncation to fit fixed context windows, raising questions about whether recall gains from fusion survive downstream constraints. These observations motivate our study: rather than evaluating fusion in isolation, we examine whether its retrieval-level benefits translate into improved evidence availability and practical gains within a production-scale RAG pipeline.

\section{Problem Statement}

Retrieval fusion is commonly adopted in Retrieval-Augmented Generation (RAG) systems under the assumption that increasing retrieval recall leads to more accurate and trustworthy answers. In enterprise environments, internal RAG systems are deployed to reduce time-to-resolution for support and operational queries, making answer accuracy critical. However, fusion also introduces additional system complexity and latency through extra retrieval passes, fusion logic, and downstream processing.

This work examines whether the added complexity and latency of retrieval fusion produce measurable improvements in answer quality for an enterprise-scale internal RAG system. In production pipelines, answer quality depends on whether relevant knowledge-base articles appear within the final Top-$K$ ranked context after reranking and truncation; improvements at the retrieval layer are therefore only meaningful if they translate into improved end-to-end evidence availability under realistic enterprise constraints.

We test the following hypotheses:

\begin{itemize}
\item \textbf{H1:} Retrieval fusion does not produce a statistically significant improvement in Top-$K$ evidence accuracy compared to a strong hybrid retrieval baseline.
\item \textbf{H2:} Recall gains introduced by fusion are largely neutralized by downstream reranking and context truncation.
\item \textbf{H3:} Any benefits from fusion are confined to recall-scarce query regimes and do not materially impact system-wide enterprise performance.
\end{itemize}

\section{Methodology}

We evaluate retrieval fusion within an enterprise-scale RAG retrieval pipeline designed to closely mirror production deployment conditions. The goal of this methodology is not to optimize retrieval performance in isolation, but to assess whether recall gains introduced by fusion persist through downstream ranking stages and improve the availability of relevant evidence under realistic enterprise constraints.

\subsection{System Baselines}

We employ a standard hybrid retrieval strategy that combines sparse lexical matching (BM25) with dense semantic retrieval, followed by a per-query cross-encoder reranking stage, reflecting common practice in large-scale information retrieval systems~\citep{robertson2009bm25}.

\begin{itemize}[leftmargin=*]
\item \textbf{Baseline}: Hybrid retrieval (BM25) followed by cross-encoder reranking (FlashRank), deduplication, and truncation to a fixed Top-$K$ context~\citep{flashrank2023}.
\end{itemize}

All baselines share the same document corpus, chunking strategy, metadata filtering rules, retrieval depth, reranker model, and ranking hyperparameters. This ensures that observed differences are attributable solely to the presence or absence of retrieval fusion. For all experiments, retrieval depth is fixed at $K=10$, and all downstream reranking, fusion, and truncation operations operate over this fixed candidate budget.

\subsection{Fusion Configuration}

We evaluate a production-faithful two-query fusion strategy. For each input query, retrieval is performed using:

\begin{itemize}[leftmargin=*]
\item \textbf{Q1}: The original user query.
\item \textbf{Q2}: A single LLM-generated paraphrase or reformulation of the original query.
\end{itemize}

For each query, a hybrid retriever returns the top $K$ document chunks. All documents are chunked using a fixed size of 512 tokens and embedded using the Granite embedding model. Query reformulations (Q2) are generated from Q1 using a LLaMA-based language model, following the prompt strategies described in \textit{~\ref{appendix:fusion_prompts}}.

Retrieval results for Q1 and Q2 are processed independently through the same cross-encoder reranker used in the baseline. Reciprocal Rank Fusion (RRF) is then applied to the two re-ranked lists to produce a fused candidate set.

Importantly, fusion is applied \emph{after} per-query reranking, reflecting production constraints where retrieval outputs are ranked independently prior to aggregation. All downstream components, including deduplication, truncation, and evaluation, remain unchanged relative to the baseline, ensuring that observed differences reflect the impact of fusion rather than embedding, chunking, or query-generation choices (see Section~6).


\subsection{Dataset and Ground Truth}

Experiments are conducted on a synthetic dataset of $N=115$ enterprise-style support queries. Each query is derived from a known knowledge-base (KB) article and paired with its originating article as ground truth, enabling controlled evaluation of retrieval behavior under realistic document distributions. Retrieval relevance is evaluated at the KB article level, while ranking operates at the chunk level using chunk-aware identifiers.

All retrieved text chunks and associated metadata are represented using standardized document abstractions provided by the LangChain framework, which is used to orchestrate retrieval and ranking components within the RAG pipeline~\citep{kakkerla2024langchain}. Product metadata is used exclusively to scope retrieval and is not used during ranking or evaluation.

\subsection{Evaluation Protocol}

End-to-end performance is evaluated using Top-$K$ KB-level hit metrics. For each query, we determine whether at least one chunk belonging to a ground-truth KB article appears within the final Top-$K$ ranked context after reranking and truncation. Metrics reported include Top-1, Top-3, and Top-10 accuracy, aggregated across the evaluation set.

This evaluation framework captures whether retrieval and fusion strategies successfully surface relevant evidence to downstream components, without relying on subjective generation quality judgments.

\subsection{Operational Constraints}

All experiments are conducted under enterprise deployment constraints that directly influence the effectiveness of retrieval fusion:

\begin{itemize}[leftmargin=*]
\item Fixed reranking capacity, limiting the number of candidates that can be meaningfully scored.
\item Context window limits, requiring truncation of ranked results prior to generation.
\item Metadata-based retrieval filtering to enforce product relevance and access control.
\end{itemize}

\section{Framework Overview}

Evaluating retrieval fusion in RAG systems requires more than comparing retrieval metrics in isolation. While fusion techniques are typically motivated by recall-oriented objectives, their downstream impact depends on how retrieved candidates interact with ranking capacity, context limits, and operational constraints within the full pipeline. The evaluation framework is designed to surface these interactions and identify where and why fusion gains may fail to translate into meaningful system-level improvements.

\begin{figure}[h]
\centering
\begin{tikzpicture}[
    node distance=9mm and 14mm,
    font=\small,
    box/.style={rectangle, rounded corners, draw, align=center, minimum width=5.0cm, minimum height=0.9cm},
    arrow/.style={-{Latex[length=2.2mm]}, thick}
]

\node[box] (qs) {Synthetic Questions (RAGAS)};
\node[box, text width=0.9\columnwidth, below=of qs] (prep) {%
Fusion Question Generation\\
(Use a prompt to generate a similar question (Q2) to the original (Q1))%
};
\node[box, below=of prep] (retr) {Retrieve Top-$K$ Chunks\\(Q1 original; optional Q2 reformulation)};
\node[box, below=of retr] (rank) {Rank Results\\(per-query re-rank; optional RRF fusion)};
\node[box, below=of rank] (eval) {Evaluate (KB-level)\\Top-1 / Top-3 / Hit@10 + Latency};
\node[box, below=of eval] (report) {Aggregate + Report\\tables, plots, trade-offs};

\draw[arrow] (qs) -- (prep);
\draw[arrow] (prep) -- (retr);
\draw[arrow] (retr) -- (rank);
\draw[arrow] (rank) -- (eval);
\draw[arrow] (eval) -- (report);

\end{tikzpicture}
\caption{Simplified overview of the evaluation pipeline.}
\label{fig:eval_flowchart_simple}
\end{figure}
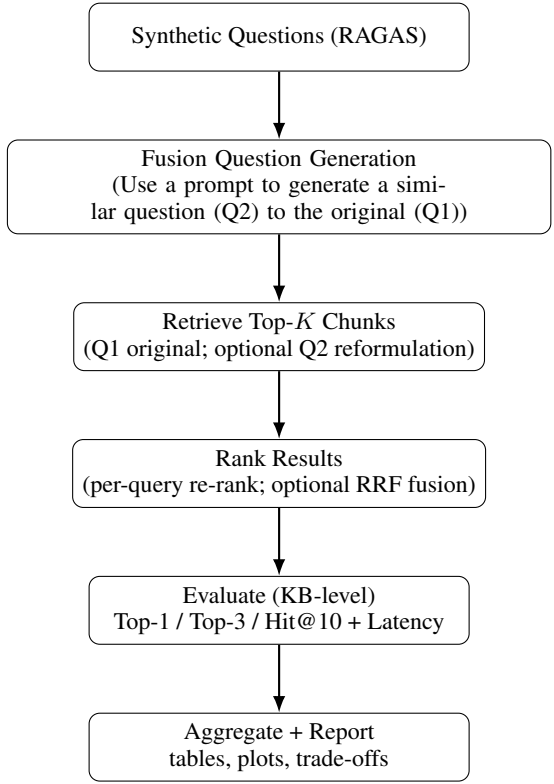

The analysis is structured around a set of recurring caveats observed when introducing fusion into production-faithful RAG pipelines, including reranker saturation, contextual redundancy, conflicting evidence, and increased operational overhead. Rather than treating these effects as secondary concerns, this work explicitly measures their impact on whether relevant evidence survives into the final ranked context.

The experimental setup follows the retrieval and ranking pipeline described in \textit{Appendix~\ref{appendix:experimental_pipeline}} and exhibited in the figure below (\textit{Figure 1}). For each query, a comparison is made using a strong hybrid retrieval baseline against a two-query fusion variant, holding all downstream components fixed. Retrieval outputs are ranked at the chunk level, fused using Reciprocal Rank Fusion, and truncated to a fixed Top-$K$ context prior to evaluation.

The primary goal is to determine whether fusion improves the availability of correct knowledge-base evidence after reranking and truncation. To this end, there is a focus on Top-$K$ KB-level hit metrics as a proxy for downstream utility. This choice allows the effect of fusion within the retrieval and ranking stages to be isolated, without conflating results with generation-specific variability.

By organizing the evaluation around concrete failure modes and measuring their impact independently, this framework enables a more precise assessment of when retrieval fusion is beneficial, when it is neutralized, and when it introduces negative trade-offs.



\section{Results}

We present results organized by the primary caveats identified in our evaluation framework. For each case, we compare the fusion configuration against the baseline under identical downstream conditions and report retrieval, ranking, and operational outcomes. Together, these results characterize how retrieval fusion behaves under realistic pipeline constraints and how its observed effects align with the hypotheses outlined in Section~3.

Tables~\ref{tab:summary_metrics_5x4_p1}--\ref{tab:summary_metrics_5x4_p3} present end-to-end accuracy for three fusion prompt strategies described in Appendix~\ref{appendix:fusion_prompts}, ranging from conservative query reformulation to explicit diversity maximization. Results are reported using KB-level Top-$K$ hit metrics, reflecting whether relevant knowledge-base evidence survives reranking and truncation under a fixed retrieval budget.

\subsection{Recall Gains Are Neutralized After Reranking}

Across all evaluated configurations, retrieval fusion consistently increases upstream recall, confirming its intended effect at the retrieval layer. However, once retrieved candidates are passed through cross-encoder reranking and truncated to a fixed Top-$K$ context, these gains are not consistently reflected in downstream KB-level accuracy. As shown in Tables~\ref{tab:summary_metrics_5x4_p1}, \ref{tab:summary_metrics_5x4_p2}, and \ref{tab:summary_metrics_5x4_p3}, fusion variants exhibit accuracy comparable to, and in some cases lower than, the baseline across Top-1, Top-3, and Hit@10 metrics.

\begin{figure}[htbp]
\centering

\begin{minipage}{\linewidth}
\centering
\begin{tabular}{lccc}
\hline
\textbf{Method} & \textbf{Top-1} & \textbf{Top-3} & \textbf{Hit} \\
\hline
baseline           & \textbf{30.43\%} & 37.39\% & \textbf{51.30\%} \\
rerank\_on\_rrf\_q1 & 29.57\% & \textbf{38.26\%} & 47.83\% \\
rerank\_on\_rrf\_2  & 15.65\% & 23.48\% & 47.83\% \\
rrf\_q1\_q2         & 25.22\% & \textbf{38.26\%} & 47.83\% \\
\hline
\end{tabular}
\captionof{table}{Summary metrics table for Fusion Prompt 1}
\label{tab:summary_metrics_5x4_p1}
\end{minipage}

\vspace{1em}

\begin{minipage}{\linewidth}
\centering
\begin{tabular}{lccc}
\hline
\textbf{Method} & \textbf{Top-1} & \textbf{Top-3} & \textbf{Hit} \\
\hline
baseline           & \textbf{29.57\%} & 36.52\% & \textbf{50.43\%} \\
rerank\_on\_rrf\_q1 & \textbf{29.57\%} & 39.13\% & 46.09\% \\
rerank\_on\_rrf\_2  & 7.83\%  & 17.39\% & 46.09\% \\
rrf\_q1\_q2         & 27.83\% & \textbf{40.87\%} & 46.09\% \\
\hline
\end{tabular}
\captionof{table}{Summary metrics table for Fusion Prompt 2}
\label{tab:summary_metrics_5x4_p2}
\end{minipage}

\vspace{1em}

\begin{minipage}{\linewidth}
\centering
\begin{tabular}{lccc}
\hline
\textbf{Method} & \textbf{Top-1} & \textbf{Top-3} & \textbf{Hit} \\
\hline
baseline           & \textbf{29.57\%} & 36.52\% & \textbf{50.43\%} \\
rerank\_on\_rrf\_q1 & \textbf{29.57\%} & 40.00\% & 44.35\% \\
rerank\_on\_rrf\_2  & 10.43\% & 19.13\% & 44.35\% \\
rrf\_q1\_q2         & 27.83\% & \textbf{40.87\%} & 44.35\% \\
\hline
\end{tabular}
\captionof{table}{Summary metrics table for Fusion Prompt 3}
\label{tab:summary_metrics_5x4_p3}
\end{minipage}

\end{figure}

These results are consistent with \textbf{H2}. Under fixed reranking and truncation constraints, additional candidates introduced by fusion frequently fail to survive into the final ranked context. While increased recall expands the candidate pool, downstream ranking capacity appears to limit the extent to which these gains can influence the evidence ultimately provided to the language model.

\subsection{Statistical Significance at Operational Top-3 Depth}
\label{sec:significance_top3}

The differences in Top-3 accuracy across fusion variants are small, so we run a paired significance test to check whether these changes are likely to be real or just noise. We treat each query as a binary hit/miss outcome: a query is a hit if any chunk from a ground-truth KB article appears in the final Top-3 ranked list after reranking and truncation (this matches the same Top-3 context we pass to the LLM during generation). Since every method is evaluated on the same $N=115$ queries, we use McNemar's exact test on paired outcomes when comparing each fusion method against the baseline (\texttt{re\_only\_q1}). Because we run multiple comparisons within each prompt type, we apply Benjamini--Hochberg correction (FDR $=0.05$). We exclude \texttt{rerank\_on\_rrf\_q2} from this analysis because it performs substantially worse than the baseline across prompts and does not represent a realistic deployment option .Table~\ref{tab:top3_sig_summary} shows that Prompt 2/3 have slightly higher Top-3 hit rates ($+2.61$ to $+4.35$ percentage points), but none of these improvements are statistically significant after correction (all $p_{\text{adj}} \ge 0.125$). In other words, fusion with certain parameters has a slightly positive impact on retrieval accuracy, but the effect is not consistent enough to confidently justify the extra fusion overhead in production.

\begin{table}[htbp]
\centering
\footnotesize
\setlength{\tabcolsep}{3.5pt}
\begin{tabular}{llcccc}
\hline
\textbf{P} & \textbf{Method} & \textbf{Top-3} & $\Delta$pp & \textbf{$p$} & \textbf{$p_{\text{adj}}$} \\
\hline
P1 & rrf\_q1\_q2         & 38.26\% & +0.87 & 1.0000 & 1.0000 \\
P1 & rerank\_on\_rrf\_q1 & 38.26\% & +0.87 & 1.0000 & 1.0000 \\
P2 & rrf\_q1\_q2         & 40.87\% & +4.35 & 0.1250 & 0.1667 \\
P2 & rerank\_on\_rrf\_q1 & 39.13\% & +2.61 & 0.2500 & 0.2500 \\
P3 & rrf\_q1\_q2         & 40.87\% & +4.35 & 0.1250 & 0.1250 \\
P3 & rerank\_on\_rrf\_q1 & 40.00\% & +3.48 & 0.1250 & 0.1250 \\
\hline
\end{tabular}
\caption{Paired Top-3 significance vs baseline.}
\label{tab:top3_sig_summary}
\end{table}

\subsection{Contextual Redundancy and Conflicting Evidence}

Fusion adds very little \emph{exact} overlap between the Q1 and Q2 retrieval lists, which means the rewrite (Q2) is usually pulling a different set of candidates than the original query. That is easy to see in Table~\ref{tab:q1_q2_overlap}: for Prompts 2 and 3 the Jaccard similarity is around $0.09$, even though the union grows to roughly 15 distinct sources. We report overlap and Jaccard because they tell us whether Q2 is actually adding new material (low overlap / low Jaccard) versus just returning the same documents in a different order, while the union size shows how much the candidate pool expands once we combine Q1 and Q2.

Low overlap by itself is not a problem---it is the point of generating a reformulation. The issue is that many of the ``new'' candidates are not truly new \emph{information}. They often repeat the same ideas (near-duplicate chunks from nearby KB sections) or introduce partially conflicting guidance across related articles. That makes reranking and truncation less stable and can push genuinely useful evidence out of the Top-$K$ window.

This also helps explain why \texttt{rerank\_on\_rrf\_q2} performs so poorly. When we fuse Q1 and Q2 with RRF and then rerank the fused list using Q2, the scoring can over-align to Q2-specific phrasing (and any drift introduced during rewriting). In practice, strong Q1-aligned documents can get demoted or dropped, which matches the sharp Top-1 drops for \texttt{rerank\_on\_rrf\_q2} in Prompts 2 and 3. Overall, fusion increases breadth, but semantic redundancy and Q2-anchored reranking can reduce the quality of the final context, aligning with \textbf{H2}.

\begin{table}[htbp]
\centering
\footnotesize
\setlength{\tabcolsep}{3.5pt}
\begin{tabular}{lcccc}
\hline
\textbf{P} & \textbf{Ov@10} & \textbf{Jac@10} & \textbf{Un@10} & \textbf{Uniq@10} \\
\hline
P1 & 2.87 & 0.279 & 13.20 & 7.82 \\
P2 & 1.09 & 0.087 & 15.29 & 7.77 \\
P3 & 1.10 & 0.094 & 14.99 & 7.77 \\
\hline
\end{tabular}
\caption{Exact Q1--Q2 overlap (Top-10).}
\label{tab:q1_q2_overlap}
\end{table}

\subsection{Limited Impact on End-to-End Evidence Availability}

Despite consistent improvements in upstream recall, fusion does not yield consistent gains in Top-$K$ KB-level accuracy across the evaluation set. In several configurations, fusion slightly degrades Top-1 and Top-3 accuracy relative to the baseline, indicating that added candidates can interfere with ranking effectiveness rather than improve it.

These findings are consistent with \textbf{H1}. Improvements at the retrieval layer do not reliably translate into better end-to-end evidence availability once downstream constraints are applied. Results are consistent across fusion prompt variants and reranking anchors within this experimental setup, suggesting that the observed behavior is not driven by a single configuration choice.

\subsection{Latency and Pipeline Overhead}

Fusion introduces additional latency due to extra retrieval passes, fusion logic, and increased downstream processing. As summarized in Table~\ref{tab:avg_latency_components}, while median latency remains within acceptable operational bounds, tail latency is consistently higher than the baseline across experimental runs. This effect compounds as retrieval depth and query volume increase.

\begin{table}[htbp]
\centering
\small
\begin{tabular}{lc}
\hline
\textbf{Metric} & \textbf{Avg Time (s)} \\
\hline
Fusion question generation & 0.89 \\
Baseline Retrieval time (Q1) & 54.60 \\
Retrieval time (Q1 + Q2 combined) & 65.98 \\

RRF time & 0.012 \\
Flashrank time ($K=10$) & 0.26 \\
\hline
\end{tabular}
\caption{Average latency components across all datapoints.}
\label{tab:avg_latency_components}
\end{table}

We observe an additional 0.89s of overhead attributable to secondary fusion processes, including fusion question generation, Reciprocal Rank Fusion, and an additional reranking pass. Retrieval latency also increases when processing two queries instead of one, reflecting the cost of additional embedding and search operations.

Although latency is not explicitly captured in the hypotheses, these results contextualize \textbf{H1} by demonstrating that fusion incurs measurable operational costs without delivering corresponding gains in evidence accuracy. In production environments where throughput and responsiveness are critical, this trade-off further constrains the practical value of fusion.

\subsection{Query-Slice Analysis}

When results are segmented by query characteristics, fusion shows modest benefits for a small subset of recall-scarce queries, particularly those for which the baseline retriever fails to surface relevant KB articles within the Top-$K$ results. In these cases, query reformulation occasionally surfaces relevant evidence that is missed by the baseline retriever.

These observations partially support \textbf{H3}. While fusion can provide localized benefits, these gains apply to a minority of the workload and do not materially affect system-wide performance. As corpus coverage and ranking capacity improve, the marginal utility of fusion diminishes, reinforcing the conclusion that its effectiveness is highly context-dependent.

\section{Conclusion}

This work presents a systematic evaluation of retrieval fusion within a production-faithful RAG retrieval and ranking pipeline. While fusion reliably improves upstream recall, our results show that these gains are frequently neutralized by downstream reranking and context constraints, and in several configurations lead to reduced evidence availability and increased latency.

By evaluating fusion through concrete pipeline bottlenecks such as reranker saturation, contextual redundancy, and operational overhead we demonstrate that recall-oriented improvements do not automatically translate into downstream utility. In mature enterprise systems, ranking capacity and context selection emerge as the dominant factors shaping end-to-end performance, often outweighing gains achieved by expanding retrieval breadth.

Our findings indicate that the benefits of fusion are highly context-dependent and largely confined to recall-scarce query regimes. For the majority of enterprise workloads, fusion introduces additional system cost without delivering material improvements in evidence availability. \textbf{While fusion may yield marginal improvements in isolated retrieval metrics, these gains are often insufficient to justify the added system complexity, computational overhead, and operational costs in production environments. The modest downstream benefits observed in our evaluation suggest that simpler, more maintainable architectures may offer superior cost-benefit tradeoffs for most enterprise RAG deployments.}

\onecolumn

\bibliography{custom}

\clearpage
\twocolumn
\appendix
\section{Experimental Pipeline and Evaluation Methodology}
\label{appendix:experimental_pipeline}

This appendix documents the end-to-end pipeline used to evaluate retrieval, fusion, and re-ranking strategies in a Retrieval-Augmented Generation (RAG) system operating over a large enterprise knowledge base. The goal here is transparency: what data went in, what intermediate artifacts were produced, how ranking was applied, and how metrics were computed. The main paper focuses on outcomes; this appendix captures the mechanics behind those outcomes.

\vspace{0.5em}
\subsection{Terminology and Core Objects}
We use the following terminology throughout the appendix.

\subsubsection{Question Set}
The \textbf{question set} is a collection of customer-support-style queries used to probe the retrieval system. In this study, questions were \textbf{synthetically generated using RAGAS}. Synthetic generation enables controlled experimentation at scale while still producing queries that resemble real support interactions~\citep{ragas2023}.

\subsubsection{Knowledge Base (KB)}
The \textbf{knowledge base (KB)} is a large text corpus containing curated domain knowledge, such as troubleshooting guides, known issues, and procedures. The KB represents subject matter expert (SME) knowledge in written form and serves as the retrieval corpus for all experiments. 

\subsubsection{KB Article and Chunks}
A \textbf{KB article} is a single document focused on a specific topic. For retrieval, each KB article is split into smaller text segments called \textbf{chunks}. Each chunk is stored as a \textbf{LangChain \texttt{Document}} object.

Chunking is used because retrieval models generally match smaller, semantically coherent spans more reliably than full-length documents, especially when articles are long or cover multiple subtopics.

\subsubsection{Index}
Our \textbf{Index} is a collection of KB articles that are embedded from text into vectors using industry standard embeddings models. The aforementioned KB article chunks are stored in this index. During the retrieval stage, the retriever performs a similarity search between the embedded query (using the same embeddings model) and the index. 
\subsubsection{Retrieval Result Document}
A \textbf{retrieval result} is a LangChain \texttt{Document} representing one retrieved chunk. Each result contains:
\begin{itemize}
    \item \texttt{page\_content}: the chunk text returned by retrieval
    \item \texttt{metadata}: fields describing the origin of the chunk, including (at minimum) its \textbf{source KB article} and \textbf{chunk index}
\end{itemize}

\subsubsection{Top-\texorpdfstring{$K$}{K} Results}
For each query, the retriever returns a ranked list of the \textbf{Top-$K$} documents (chunks). K is a value defined by us, in this case K=10. When results from multiple queries are logged in a single list, we keep a fixed positional convention:

\begin{table}[h]
\centering
\begin{tabular}{ll}
\hline
Positions & Meaning \\
\hline
$1 \ldots K$ & Q1 retrieval results \\
$K{+}1 \ldots 2K$ & Q2 retrieval results \\
\hline
\end{tabular}
\caption{K Retrieval Results}
\label{tab:k_retrieval_positions}
\end{table}

\vspace{0.5em}
\subsection{Dataset Construction and Ground Truth}

\subsubsection{Query Dataset}
Each experimental run operates over a dataset where each row corresponds to a single query instance. At minimum, each row contains:
\begin{itemize}

    \item \texttt{question}: the natural-language query text (synthetically generated via RAGAS)
    \item \texttt{ground\_truth}: the 'true' KB article from which the question was generated (Section A.2.2)
\end{itemize}

In practice, additional metadata may be present (e.g., product category labels, query length). This metadata is used only for filtering and corpus scoping (Section~\ref{appendix:filtering}), and is not directly used as a ranking signal.

\subsubsection{Ground Truth Labels}
Relevance is defined at the \textbf{KB article level}. Each \texttt{question} is associated with one or more valid 'ground truth' KB articles:
\begin{itemize}
    \item \texttt{kb\_article\_id}: one or more KB article identifiers considered correct for that query 
\end{itemize}

During evaluation, a retrieval is considered a \textbf{hit} if \emph{any} ground-truth KB article is retrieved (via any of its chunks) within the top-$K$ ranked results.

\vspace{0.5em}
\subsection{Document Representation}
\label{appendix:doc_representation}

\subsubsection{Chunking Strategy}
All KB articles are preprocessed into chunks prior to indexing. Each chunk is represented as a LangChain \texttt{Document} with \texttt{page\_content} (text) and \texttt{metadata}. The metadata includes:
\begin{itemize}
    \item \texttt{source\_kb}: identifier of the KB article the chunk came from
    \item \texttt{chunk\_index}: the chunk position within the article
    \item optional fields used for filtering and access control (if applicable)
\end{itemize}

\subsubsection{Chunk-Aware Identifiers}
Because multiple chunks can come from the same KB article, we treat the chunk as the atomic retrieval unit. To avoid accidentally collapsing multiple relevant chunks into a single item during fusion or re-ranking, each chunk is assigned a unique identifier:
\begin{table}[h]
\centering
\small
\begin{tabular}{l@{\hspace{1em}}l}
\hline
\textbf{Format} & \texttt{source\_kb::chunk=chunk\_index} \\
\textbf{Example} & \texttt{KB123456::chunk=0} \\
\hline
\end{tabular}
\caption{Source identifier format for KB chunk references.}
\label{tab:kb_chunk_id_format}
\end{table}

This chunk-aware identifier is carried through retrieval, fusion, and re-ranking. Importantly, it is normal for multiple chunks from the same KB article to appear in the same top-$K$ list. We want to maintain a chunks position during re-ranking to ensure that if multiple chunks of the same KB article are used to formulate an answer we don't lose any of that information.

\vspace{0.5em}
\subsection{Retrieval Stage}

\subsubsection{Metadata Filtering}
\label{appendix:filtering}
A lightweight metadata filtering layer is used to scope retrieval to a relevant portion of the KB. The purpose is operational rather than theoretical: limiting the candidate set can improve precision and latency by excluding articles that are clearly outside the query's target product category (or other high-level grouping).

Filtering is applied \emph{before} retrieval (i.e., it constrains the searchable subset), and does not directly change scoring or ranking logic.

\subsubsection{Two-Query Retrieval for Fusion Experiments}
To evaluate fusion strategies, we retrieve using two query formulations:
\begin{itemize}
    \item \textbf{Original Query (Q1)}: the original question text
    \item \textbf{Fusion Query (Q2)}: an LLM-generated reformulation (e.g., paraphrase) intended to surface alternate matching evidence
\end{itemize}

For each query ($Q1$ and optionally $Q2$), the retriever returns top-$K$ chunks as LangChain \texttt{Document} objects. The ordered outputs are concatenated using the positional convention shown earlier.

\subsubsection{Fusion Question Generation Prompts}
\label{appendix:fusion_prompts}
In order to better address our problem statement, we decided to perform the experiment with three different fusion generation prompts. These prompts are sent to the LLM as instructions along with the original question. The LLM then generates a similar question using re-wording strategy defined in the instruction, as well as some information about the question (product category, etc).

\begin{figure}[H]
    \centering
    \fbox{%
        \begin{minipage}{0.90\linewidth}
            \small\ttfamily
            <|instruction|>\\
            You are a technical query rewriter.\\
            You will be given a technical query, case details and product information.\\
            Re-write the query without adding details about the product or case.\\[4pt]
            Guidelines:\\
            - Simply give the output, not your reasoning for it\\
            - Don't include any extra details other than the ones provided in the input\\
            - Keep it Direct and Concise\\
            </instruction>
        \end{minipage}%
    }
    \caption{Prompt Type 1}
    \label{fig:fusion_prompt_1}
\end{figure}

\begin{figure}[H]
    \centering
    \fbox{%
        \begin{minipage}{0.90\linewidth}
            \small\ttfamily
            <|instruction|>\\
            You are a technical query rewriter.\\
            Your role is to re-write a given technical query for  Troubleshooting.\\
            Utilize the given information to re-write the technical query such that it is optimized for a RAG system.\\[4pt]
            Guidelines:\\
            - Simply give the output, not your reasoning for it\\
            - Don't include any extra details other than the ones provided in the input\\
            - Make the query precise and detail-oriented without making it too long.\\
            </instruction>
        \end{minipage}%
    }
    \caption{Prompt Type 2}
    \label{fig:fusion_prompt_2}
\end{figure}

\begin{figure}[H]
    \centering
    \fbox{%
        \begin{minipage}{0.90\linewidth}
            \small\ttfamily
            <|instruction|>\\
            You are a retrieval strategist.\\
            Your job is to generate diverse, high-quality rewrites of a technical query to maximize document recall in a RAG fusion pipeline.\\
            You must only return ONE high-quality re-write.\\[4pt]
            Guidelines:\\
            - Simply give the output, not your reasoning for it\\
            - Don't include any extra details other than the ones provided in the input\\
            </instruction>
        \end{minipage}%
    }
    \caption{Prompt Type 3}
    \label{fig:fusion_prompt_3}
\end{figure}

\vspace{0.5em}
\subsection{Re-Ranking and Fusion Pipeline}
The experimental pipeline evaluates a production-like ranking stack composed of multiple stages. Not every stage is used in every configuration, but the outputs are logged consistently to allow direct comparison.

\subsubsection{Per-Query Re-Ranking (Baseline)}
To emulate an enterprise setup where retrieval outputs are refined before any fusion is applied, each query's retrieved list is re-ranked independently using the flashrank module. This yields:
\begin{itemize}
    \item \texttt{baseline}: re-ranked list for Q1 retrieval results 
    \item \texttt{q2\_baseline}: re-ranked list for Q2 retrieval results 
\end{itemize}

Each list is stored as an ordered sequence of chunk-aware identifiers. If multiple chunks from the same KB article are ranked highly, they are preserved. The re-ranked list of Q1's retrieval results is our \textbf{baseline} datapoint. Refer to Tables 8,9 and 10 below.

\subsubsection{Reciprocal Rank Fusion (RRF)}
Reciprocal Rank Fusion (RRF) is applied to combine multiple ranked lists into a single fused ranking. In this study, RRF is applied over the already re-ranked lists (\texttt{baseline}, \texttt{q2\_baseline})

For each ranked list $L_i$, a chunk $d$ at rank $r$ contributes:
\[
\text{score}(d) \;{+}{=}\; \frac{1}{k + r}
\]
where $k$ is a smoothing constant.

Key properties of the implementation:
\begin{itemize}
    \item Fusion is performed at the \textbf{chunk level} using chunk-aware identifiers.
    \item Each chunk contributes at most once per list (best rank is used if duplicates occur).
    \item The output is a single fused ordered list, denoted \texttt{rrf\_q1\_q2}.
\end{itemize}

\subsubsection{Post-Fusion Re-Ranking}
To isolate whether fusion benefits from additional semantic refinement, we optionally re-rank the top-$K$ items from the fused list using flashrank. This produces two variants, depending on which query formulation is used as the re-ranking anchor:
\begin{itemize}
    \item \texttt{rerank\_on\_rrf\_q1}: re-rank fused list using Q1 as the anchor
    \item \texttt{rerank\_on\_rrf\_q2}: re-rank fused list using Q2 as the anchor
\end{itemize}

This stage helps separate:
\begin{itemize}
    \item the effect of \textbf{fusion alone}, versus
    \item the effect of \textbf{fusion + additional re-ranking}.
\end{itemize}

To see the evaluation of these lists, refer to Tables 8, 9 \& 10 below.

\vspace{0.5em}
\subsection{Recorded Outputs}
For each instance of a pair of queries (Q1, Q2)  (\texttt{question}), the experiment logs:
\begin{itemize}
    \item ranked output lists for each pipeline configuration (per-query re-ranks, RRF fused list, post-fusion re-ranks)
    \item retrieval latency measurements
    \item re-ranking latency measurements
    \item fusion question generation latency measurements (when enabled)
\end{itemize}

All ranked lists are stored as ordered sequences of chunk-aware identifiers, enabling consistent evaluation even when chunk text or metadata changes over time.

\vspace{0.5em}
\subsection{Evaluation Methodology}

\subsubsection{KB-Level Evaluation from Chunk-Level Rankings}
Although retrieval and ranking operate on chunks, correctness is defined at the KB article level. To evaluate results, we map each chunk-aware identifier back to its KB article by removing the chunk suffix:
\[
\texttt{source\_kb::chunk=idx} \;\rightarrow\; \texttt{source\_kb}
\]

A method is counted as a \textbf{hit} for a query if \emph{any} chunk belonging to \emph{any} ground-truth KB article appears within the top-$K$ ranked results for that method.

This design matches how a RAG based conversational model works. Retrieving any strong supporting passage from the correct article is usually sufficient for downstream answer generation, even if the exact chunk differs.

\subsubsection{Metrics}
For each ranking method, we compute:
\begin{itemize}
    \item \textbf{Top-1 accuracy}: fraction of queries where a ground-truth KB article appears at rank 1
    \item \textbf{Top-3 accuracy}: fraction of queries where a ground-truth KB article appears within ranks 1--3
    \item \textbf{Hit@10 (Top-10 hit rate)}: fraction of queries where a ground-truth KB article appears within ranks 1--10
\end{itemize}

Metrics are computed per query and then aggregated across the dataset.

\subsubsection{Evaluation Artifacts}
The primary evaluation artifact is a \textbf{summary metrics table} with one row per method and columns for Top-1, Top-3, and Hit@10. This is the table used in the main paper for reporting and comparison.



\begin{table}[htbp]
\centering
\begin{subtable}{\linewidth}
\centering
\begin{tabular}{lccc}
\hline
\textbf{Method} & \textbf{Top-1} & \textbf{Top-3} & \textbf{Hit} \\
\hline
baseline           & 0.3043 & 0.3739 & 0.5130 \\
rerank\_on\_rrf\_q1 & 0.2957 & 0.3826 & 0.4783 \\
rerank\_on\_rrf\_2  & 0.1565 & 0.2348 & 0.4783 \\
rrf\_q1\_q2         & 0.2522 & 0.3826 & 0.4783 \\
\hline
\end{tabular}
\caption{Summary metrics table for Fusion Prompt 1}
\end{subtable}

\vspace{1em}

\begin{subtable}{\linewidth}
\centering
\begin{tabular}{lccc}
\hline
\textbf{Method} & \textbf{Top-1} & \textbf{Top-3} & \textbf{Hit} \\
\hline
baseline           & 0.2957 & 0.3652 & 0.5043 \\
rerank\_on\_rrf\_q1 & 0.2957 & 0.3913 & 0.4609 \\
rerank\_on\_rrf\_2  & 0.0783 & 0.1739 & 0.4609 \\
rrf\_q1\_q2         & 0.2783 & 0.4087 & 0.4609 \\
\hline
\end{tabular}
\caption{Summary metrics table for Fusion Prompt 2}
\end{subtable}

\vspace{1em}

\begin{subtable}{\linewidth}
\centering
\begin{tabular}{lccc}
\hline
\textbf{Method} & \textbf{Top-1} & \textbf{Top-3} & \textbf{Hit} \\
\hline
baseline           & 0.2957 & 0.3652 & 0.5043 \\
rerank\_on\_rrf\_q1 & 0.2957 & 0.4 & 0.4435 \\
rerank\_on\_rrf\_2  & 0.1043 & 0.1913 & 0.4435 \\
rrf\_q1\_q2         & 0.2783 & 0.4087 & 0.4435 \\
\hline
\end{tabular}
\caption{Summary metrics table for Fusion Prompt 3}
\end{subtable}

\label{tab:summary_metrics_all}
\end{table}

\subsection{Robustness and Failure Handling}
We took steps to ensure that our iterative process is controlled and consistent. To maintain experimental integrity and avoid mixing partial outputs:
\begin{itemize}
    \item retrieval calls are retried on transient failures (e.g., timeouts or backend instability)
    \item queries that fail after the maximum retry budget are skipped and logged
    \item downstream stages (re-ranking and fusion) assume retrieval succeeded, preventing partially populated lists from contaminating metrics
\end{itemize}

\subsection{Reproducibility}
Given fixed retriever parameters, fixed re-ranker models, and fixed fusion hyperparameters, the pipeline is deterministic. Chunk-aware identifiers provide a stable unit of comparison across stages and prevent accidental collapse of evidence when multiple chunks originate from the same KB article.

\end{document}